\documentclass[aip,ajp,prb,reprint,groupedaddress]{revtex4-1} 
 \pdfoutput=1

\usepackage{amsmath}  
\usepackage{amsfonts} 
\usepackage{graphicx} 

\usepackage{times}
\usepackage{graphicx,epsfig}

\begin{document}

\title{Vision and change in introductory physics for the life sciences}

\author{S. G. J. Mochrie}
\affiliation{Departments of Physics and Applied Physics, Yale University, New Haven, Connecticut 06511}


\begin{abstract}
Since 2010, the Yale physics department has offered a novel
calculus-based introductory physics for the life science (IPLS) sequence,
that re-imagines the IPLS syllabus  to include a selection of biologically and medically relevant topics, that are highly meaningful to its audience of biological science and premedical undergraduates.
The first semester, in particular, differs considerably from traditional
first-semester introductory physics.
Here, we highlight the novel aspects of Yale's first-semester course,
and describe student feedback about the course, including
a comparison between how students
evaluate the course and  how they evaluate courses with a traditional syllabus,
and how students' perceptions of the relevance of physics to biology and medicine
are affected by having taken the course.
\end{abstract}

\maketitle

\section{Introduction}

\begin{quote}
``To design something really new and innovative you have to reject reason.''
Jonathan Ive.
\end{quote}

When biology and premedical students arrive in introductory physics courses,
they are often skeptical about the relevance of physics and mathematics to
their academic and professional goals.
Life science faculty are also
increasingly  questioning the value of  traditional introductory physics
in their majors'
crowded schedules.
For example, in the Fall 2014 issue of the American Physical Society's
Forum on Education,
University of Colorado
molecular, cellular, and developmental biology professor Michael Klymkowsky writes
\cite{Klymkowsky2014}:
\begin{quote}
``These ... physics courses are rarely designed to meet the needs of biology students, and in many cases, little thought has gone into articulating exactly why students should be required to take them.''
\end{quote} 
He continues:
\begin{quote}
``I would reject the premise that physics {\em per se}
 is generically useful to understanding molecular biology.
A poorly designed course,
perceived as irrelevant to the disciplinary interests or
needs of students could be viewed as an inappropriate imposition.''
\end{quote}

However, in contrast to the perception that the traditional introductory physics
curriculum may not be meeting the needs of today's undergraduate biology students,
the value of physics ideas and physics skills in biology
are becoming ever more appreciated,
reflecting the fact that biology
is experiencing an ongoing transformation into a quantitative science.
More than ten years ago, this evolution, which has since continued apace,
was eloquently described  by Princeton
physics professor
William Bialek
and Princeton molecular biology professor David Botstein in a seminal paper
\cite{BialekBotstein}:
\begin{quote}
``Dramatic advances in biological understanding, coupled with equally dramatic advances in experimental techniques and computational analyses, are transforming the science of biology. The emergence of new frontiers of research in functional genomics, molecular evolution, intracellular and dynamic imaging, systems neuroscience, complex diseases, and the system-level integration of signal-transduction and regulatory mechanisms require an ever-larger fraction of biologists to confront deeply quantitative issues that connect to ideas from the more mathematical sciences.
At the same time, increasing numbers of physical scientists and engineers are recognizing that exciting frontiers of their own disciplines lie in the study of biological phenomena. Characteristic of this new intellectual landscape is the need for strong interaction across traditional disciplinary boundaries.''
\end{quote}
In response to this new landscape,
biology departments are now increasingly hiring faculty with physics backgrounds,
and biological physics is
now a major subfield of physics,
well represented in physics departments and professional societies across the world.

At the undergraduate level,
a number of reports  \cite{BIO2010,AAMC,VCUBE,Convergence2014}
have highlighted the increasing importance of quantitative skills
for students who are planning biomedical careers,
and the need to modify and augment undergraduate
biology and premedical education accordingly.
Recently,  the biology community,
in {\em Vision and Change in Undergraduate Biology Education} (VCUBE),
  has specified a number of core competencies that all
undergraduate biology students should possess  \cite{VCUBE}.
These competencies and examples of how each competency
might be demonstrated in practice
are summarized in VCUBE's Table 2.1, reproduced here as Table~\ref{VCUBEMatrix}.
The AAMC/HHMI's {\em Scientific Foundations for Future Physicians} (SFFP)  specifies similar competencies
for premedical students \cite{AAMC}.


Inspection of Table~\ref{VCUBEMatrix}
reveals that undergraduate biology education in the twenty-first century
must embrace quantitative and mathematical approaches.
Remarkably, many of the specified competencies
are those that physicists seek for students to acquire in physics classes.
In this context, the two-semester introductory physics for the life sciences (IPLS) sequence,
currently required for premedical students and biological science majors,
presents a natural platform
where these students could encounter quantitative and mathematical descriptions
of biological and physiological phenomena for the first time.
Where better than IPLS
to first develop
problem-solving strategies?
Where better than IPLS 
to first develop
the ability to use quantitative reasoning?
Where better than IPLS 
to start developing
the ability to use modeling and simulation?
Where better than IPLS
to start developing
the  ability to apply physical laws to biological dynamics?
Where better than IPLS
to start developing
the ability to incorporate stochasticity into biological models?
Indeed, VCUBE  can be read
as a call to transform IPLS into
an engaging and exciting subject that is appreciated as essential
to every biologist's undergraduate education,
in  contrast to biology students'  current preconceptions.

\begin{table*}[t!]
\center
\begin{tabular}{|c||c|c|c|c|c|c|}
\hline
Core&Ability to apply&{Abilty to use}&Ability to use&Ability to&Ability to&Ability to\\
Competency&the process of&quantitative&modeling and&tap into the&communicate&understand\\
&science&reasoning&simulation&interdisciplinary&and&the\\
&&&&nature of&collaborate&relationship\\
&&&&science&with other&between\\
&&&&&disciplines&science and\\
&&&&&&society\\
\hline
\hline
 Instantiation&Biology is an&Biology relies  &Biology&Biology is an  &Biology is a&Biology is\\
of ability in& evidence-based&on applications&focuses&interdisciplinary&collaborative&conducted\\
disciplinary&discipline& of quantitative&on the study&science&scientific&in a societal\\
practice&&analysis and&of complex&&discipline&context\\
&&mathematical&systems&&&\\
&&reasoning&&&&\\
\hline
Demonstration&Design scientific&Apply&Use&Apply concepts&Communicate&Identify\\
of competency&process to&quantitative&mathematical&from other&biological&social and\\
in practice&understand&analysis to&modeling and&sciences to&concepts and&historial\\
&living systems&interpret&simulation&interpret&interpretations&dimensions\\
&&biological&tools to&biological&to scientists&of biology\\
&&data&describe&phenomena&in other&practice\\
&&&living systems&&disciplines&\\
\hline
Examples&Observational&Developing and&Computational&Applying physical&Scientific&Evaluating\\
of core&strategies&interpreting&modeling of&laws to biological&writing&the\\
competencies&&graphs&dynamic&dynamics&&relevance\\
applied&Hypothesis&&systems&&Explaining&of social\\
to biology&testing&Applying&&Chemistry of&scientific&contexts to\\
practice&&statistical&Applying&molecules and&concepts&biological\\
&Experimental&methods to&informatics&biological&to different&problems\\
&design&diverse data&tools&systems&audiences&\\
&&&&&&Developing\\
&Evaluation of&Mathematical&Managing and&Applying&Team&biological\\
&experimental&modeling&analyzing large&imaging&participation&applications\\
&evidence&&data sets&technologies&&to solve\\
&&Managing and&&&Collaborating&societal\\
&&analyzing large&Incorporating&&across&problems\\
&Developing&data sets&stochasticity&&across&\\
&problem-solving&&into biological&&disciplines&Evaluating\\
&strategies&&models&&&ethical\\
&&&&&Cross-cultural&implications\\
&&&&&awareness&of biological\\
&&&&&&research\\
\hline
\end{tabular}
\caption{Table 2.1 of the AAAS's {\em Vision and change in undergraduate biology education} (VCUBE).
}
\label{VCUBEMatrix}
\end{table*}

\begin{figure}[t!]
\centering
{\includegraphics[width=0.48\textwidth,keepaspectratio=true]{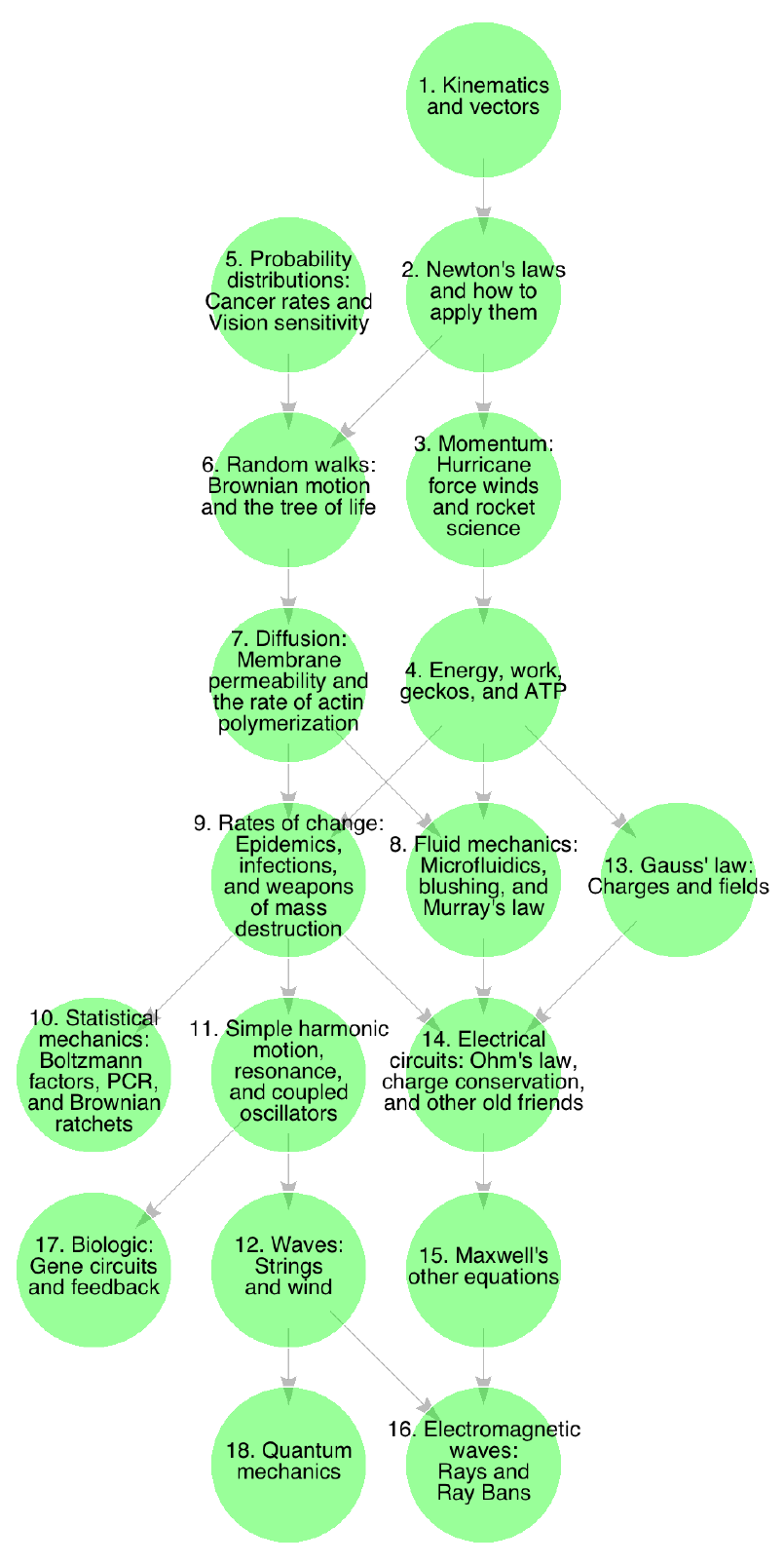}}
\caption{Syllabus for PHYS 170 {\em University Physics for the Life Sciences},
represented as a concept map.
Each module is depicted as a green circle.
A module receiving an arrow indicates that the material in that module
depends on the material in
the module sending the arrow and preceding modules. 
Modules 1 to 10 constitute the material for the first semester.
Modules 11 to 18 constitute the material for the second semester.
}
\label{OverallConceptMap}
\end{figure}

These considerations have lead to considerable recent interest in
IPLS reform, including the approaches
described in Refs. \onlinecite{Meredith2012,Redish2014,Crouch2014,IPLSConferenceReport}.
For the last six years, the Yale physics department has offered a calculus-based IPLS sequence
 -- PHYS 170/171 {\em University Physics for the Life Sciences}  -- 
that re-imagines the introductory physics syllabus to
include
a selection of biologically
and medically relevant topics,
that are highly meaningful to the intended audience,
and that permit a number of the competencies specified in VCUBE to be addressed.
The format of PHYS 170 is two 75-minute class sessions per week for all $\sim$140 students in the class,
50-minute discussion sections,
each for about twenty students, lead by one of the course graduate-student teaching assistants,
and optional ``study halls'', staffed by course personal to provide guidance,
two evenings each week, where students
work together on the course problem sets.
The laboratory component of introductory physics at Yale
is a separate (half-credit) class that continues to follow a more traditional syllabus.
For the first three years, PHYS 170 was offered as a traditional lecture class.
More recently, it has used a TEAL-format classroom \cite{Dori2005}, and has required more active participation of the students.

In the present manuscript, first,  I briefly describe how the syllabus
of the first semester, PHYS 170,
differs from that of traditional first-semester introductory physics.
The main goal of this paper, however,
is to describe how PHYS 170 has been received by its audience,
how that reception differs from the corresponding receptions of 
Yale's traditional first-semester introductory physics
offerings, and how students'
perceptions of the relevance of physics to biology and medicine
are affected by the course.

\section{Introductory physics re-imagined}
The overarching goals for PHYS 170 are as follows:
\begin{itemize}

\item Introduce biological science majors and future clinicians to
physical  and mathematical principles and tools,
that will enable a deeper scientific understanding of biological systems,
including the human body, and how they may be studied or diagnosed.

\item Demonstrate the application of physics and mathematics to the life sciences
and medicine and the human
body, via relevant and authentic examples.

\item Seed an enduring appreciation of the power of physical and mathematical
approaches in biology and medicine, and in science more broadly.

\item Transform introductory physics for the life sciences into an engaging and
exciting subject, that is understood to be essential to every biologist's undergraduate education
by biology students and faculty alike.

\item Develop curricula materials, that will enable introductory
physics instructors to teach a biologically-relevant IPLS at their own institutions.
%
\end{itemize}
An overview of the PHYS170/171 syllabus
is shown in  Fig. ~\ref{OverallConceptMap},
represented as a concept map.
The course consists of 18 modules, each depicted as a green circle.
Modules 1 to 10 constitute the material for the first semester (PHYS 170).
Modules 11 to 18 constitute the material for the second semester (PHYS 171).
Modules 1-4, 11 and 12 constitute a fairly ordinary treatment of vectors,
kinematics, Newtonian mechanics
(excluding rotational dynamics),
energy conservation, simple harmonic motion, and wave motion.
Nevertheless, these modules do include a number of
biologically-authentic examples, such as
Ramachandran plots \cite{Ramachandran1963},
tensegrities \cite{Ingber2014},
biofilament buckling,
and how geckos can walk up (and not just stick to) walls.
Module 13 covers electrostatics, focussing on Gauss' law, but it includes a section on
screened electrostatic interactions, which are pervasive in biology.
Module 14 treats electrical circuits,
but relies on analogies to diffusive and fluid transport
that students have previously encountered in modules 7 and 9, respectively.
Modules 15 and 16 discuss Maxwell's equation and electromagnetic waves.
This material is included because
 Maxwell's equations continue to set the standard
for what physicists consider to be beautiful and elegant.
In addition,  electromagnetic wave-based studies,
such as optical microscopy and x-ray crystallography,
are tremendously important in biomedical research.
There is a short module on quantum mechanics (module 18), in part because of
quantum mechanics' scientific importance, but also at the urging of a number of
faculty colleagues in Yale's biological science departments.

Modules 5-10 and 17, however, deviate significantly from a traditional syllabus.
In these modules, we 
discuss probability (module 5),
random walks (module 6),
diffusion (module 7),
low Reynolds  number fluid flow (module 8), 
mathematical modeling (module 9),
 statistical mechanics, applied to a number of molecular-biologically important processes (module 10),
 and gene circuits and feedback (module 17).
 Modules 10 and 17 are
 described in detail in Ref.~\onlinecite{Mochrie2011}
 and Ref.~\onlinecite{Cahn2014}, respectively.

In developing this syllabus, a number
of recent excellent advanced-level textbooks at the interface of physics and biology
have been inspirational and  tremendously useful
\cite{BergBook,Nowak2000,Nelson,Nowak2006,PhillipsKondevTheriot}.
The syllabus was also influenced
by Refs.~\onlinecite{BIO2010,AAMC,VCUBE,Convergence2014}, and by
discussions with colleagues within science and engineering departments across Yale.
A key consideration was
the fact that this will likely  be the final physics course that its students will take, and that therefore
there is no rationale to prepare students for more advanced physics classes, nor any excuse to
postpone compelling material for a later course.
Overall, this first-semester syllabus is the result of a backwards design process \cite{Handelsman2007a}
that started with a number of key ideas and topics  --
diffusion, fluid flow, mathematical modeling, and statistical mechanics --
that it would be necessary to include to achieve the overarching goals listed above.

Module 6  on random walks is included to introduce students to 
 the ubiquitous role of  stochastic  processes in biology.
Our first random walk example is Brownian motion. We point out
that Einstein's theory of Brownian motion and its subsequent confirmation
by Perrin were pivotal in finally convincing skeptics at the outset of the twentieth
century that atoms and molecules really exist.
We also discuss that, in evolutionary biology,  random walks have
a central role in evolution via ``genetic drift''.
Genetic drift is the random process -- it can be conceived as a random walk
 -- by which ``neutral'' mutations,
that offer no selective (dis)advantage,
become fixed in a population,
leading the population's genomes to evolve, or drift, away from their initial state,
over many generations, as such neutral mutations accumulate.
Analyses of genetic drift in order to establish evolutionary
relationships between organisms and among populations of a single organism \cite{Cann1987},
represent a pivotal development in evolutionary biology.
The theoretical underpinnings of such DNA-based phylogenetic analyses
critically rely on the properties of  genetic drift's random walk.

Module 7 concerns diffusion, which is tremendously important across the sciences and in medicine.
We show how diffusion is a direct manifestation of individual particles' random walks.
Specifically, we focus on diffusion across membranes,
which is particularly important in physiology and medicine,
 and  the ``diffusion to capture'' of actin monomers
 at the tip of an actin filament, {\em i.e.} actin polymerization.
  Actin polymerization is an important example of biological dynamics, 
 underlying many examples of eukaryotic cell motility.
More generally, our discussion of diffusion to capture reveals how
essential equations of chemical
and biochemical kinetics emerge from particles'  Brownian motion
and their consequent diffusion.
On the more biological side,
diffusion to capture also permits us to discuss the physics constraints that require
that the sizes of individual animal cells,
from those of roundworms (1~mm long) to those of blue whales (30~m long),
are always about the same size (10~$\mu$m), even for animals differing in volume by $10^9$ or so.

Module 8 on fluid mechanics focusses on
laminar fluid flow in microfluidic devices
and especially on the flow of blood through the circulatory system.
We discuss atherosclorosis and how you blush.
We also discuss fluid flows in fluid circuits,
which is closely analogous to current flows in electrical circuits.
The module concludes with a discussion of the principles that underlie
the physiology of the human circulatory system (Murray's Law),
which demonstrates how the physics of
viscous liquid friction plausibly has determined human physiology
\cite{Murray1926}.

Module 9 provides a final example that
stands outside usual introductory physics.  This module
focuses on how a number of 
interesting processes can be modeled 
mathematically.
Several  disparate processes are described by similar equations.
We are thus able to point out that the solutions and understanding gained
in one case can be transferred to another.
When colleagues first hear our topics,
some complain: ``that's not physics.''
Our riposte is that
the equations that describe the early time course of
HIV viral load in an individual patient following infection can
be re-interpreted to describe an atomic explosion; and
the equations that describe
the occurrence of retinoblastoma --
the most common childhood eye cancer -- that results from somatic mutations
are formally identical to equations that describe
the number of a particular species of radioactive nuclei in
a certain radioactive decay chain.
We explain these applications and point out their
connexions in the course of the module.
The medical {\em bona fides} of these examples are compelling:
The mathematical modeling of HIV progression in an individual
was important in the development
of HIV treatments \cite{Wei1995,Ho1995,Perelson1996};
and the age-dependence of retinoblastoma onset lead to the
hypothesis of tumor suppressor genes,
which proved a tremendously important step in understanding cancer \cite{Knudson1971}.
This module also well positions the class to discuss
genetic circuits in the second semester \cite{Cahn2014}.

 \begin{figure*}[t!]
\centering
{\includegraphics[width=0.95\textwidth,keepaspectratio=true]{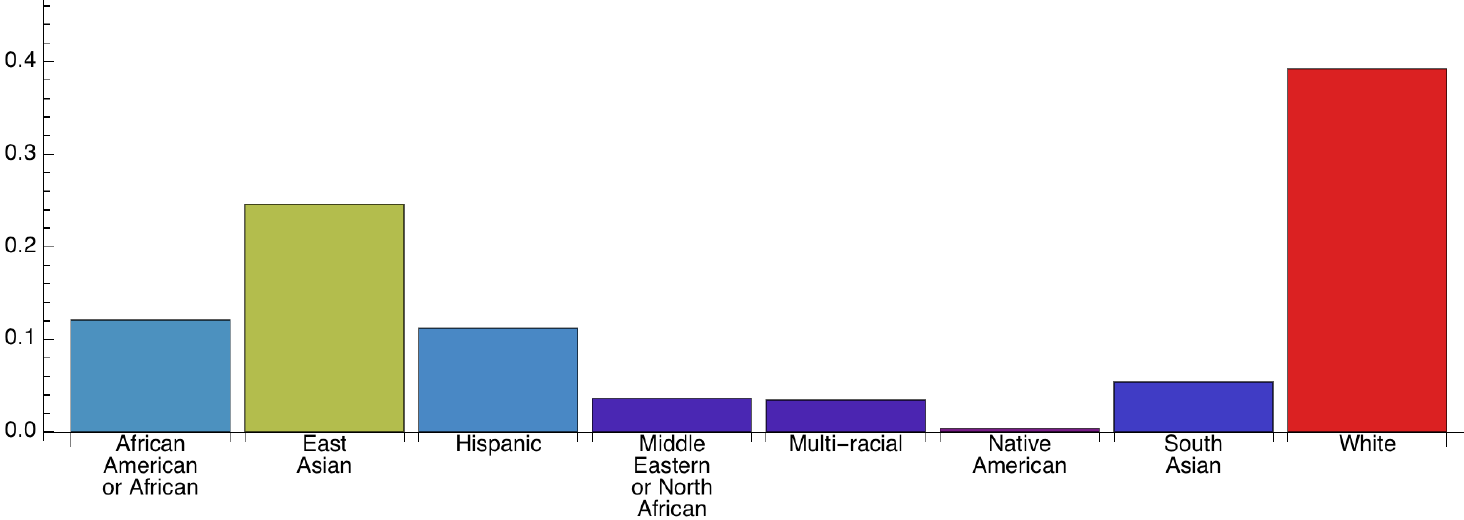}}
\caption{
Ethnicities of
{\em Physics for the Life Sciences} students,
specified as a fraction of the total number of respondents.
}
\label{Ethnicity}
\end{figure*}

\section{Audience}
%
%

\begin{figure}[b!]
\centering
{\includegraphics[width=0.45\textwidth,keepaspectratio=true]{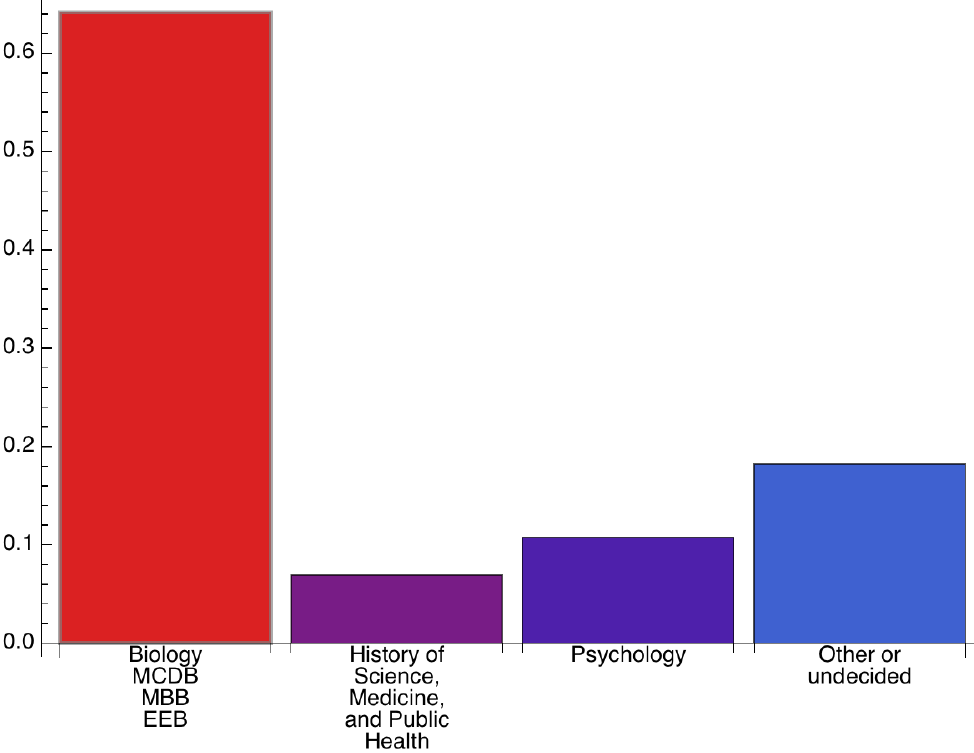}}
\caption{
Planned major  of
{\em Physics for the Life Sciences} students,
specified as a fraction of the total number of respondents.
MCDB=Molecular, Cellular, and Developmental Biology;
MBB=Molecular Biophysics and Biochemistry;
EEB=Ecology and Evolutionary Biology.
}
\label{PlannedMajor}
\end{figure}

Since 2010, over 700 students have taken PHYS 170;
66\% are female; 34\% are male.
They are ethnically diverse (Fig. ~\ref{Ethnicity}),
 including significant numbers of students from groups underrepresented in STEM
 disciplines ({\em e.g.} 10\% African American or African and 10\% Hispanic). 
 The overwhelming majority of these students will also have taken Yale's introductory chemistry
and biology sequences, before arriving in PHYS 170,
and therefore they
possess considerable chemical and biological sophistication. 
Their overall mathematical skills, however, are rusty,
but improve tremendously over the course.
64\% are biological science majors.
There are also significant numbers of Psychology majors
and History of Science, Medicine, and Public Health majors
(Fig. ~\ref{PlannedMajor}).
More than 80\%  identify themselves as premedical students.
In many cases, they are involved in biological or biomedical research,
and in medically-related volunteer work.
After graduation, many go on to highly-ranked medical and graduate schools.
All of them take the class
in order to fulfill the physics requirement for medical school and/or
the physics requirement of their major.

\section{Mathematics}
\begin{quote}
``I have deeply regretted that I did not proceed far enough at least to understand
something of the great leading principles of mathematics, for men thus endowed seem to have an extra sense.''  Charles Darwin.
\end{quote}
Because all students in PHYS 170 would
have previously taken a first course in calculus,
our starting point was a calculus-based class.
We also decided to emphasize the use of Wolfram Alpha \cite{WolframAlpha},
which provides a 
 web browser interface, that facilitates mathematical manipulations of all sorts.
Using Wolfram Alpha empowers students to carry out more
sophisticated mathematics than otherwise.
Early on in the semester, we offer a topical primer on how to use Wolfram Alpha.
However, students are lead to really engage with Wolfram Alpha
 by at least one homework problem each week that requires its use.
We also include  simulations and calculations, via a number of Wolfram Demonstrations,
which students can conveniently run in their web browsers and interact with via sliders.
To be clear, no knowledge of Mathematica is needed of the students.

The mathematical style and level of the course was  informed by the view that
 that twenty-first-century biology
 will be well-served by a version of IPLS  that seeks to contribute to
closing any mathematical and quantitative gaps between physical and life science students.
This approach stands apart from other recent IPLS courses
 \cite{Meredith2012,Redish2014,Crouch2014},
 which also significantly modify the traditional syllabus,
 but which tend to work within the historical difference in culture
between the physical sciences and the life sciences,
namely that the life sciences have generally not embraced
mathematics and quantitative analyses,
 in contrast to the physical sciences.

Beyond calculus, the mathematics that we employ is dictated by the topics that we seek to treat,
{\em i.e} we also used backwards design here.
To discuss Brownian motion and statistical mechanics, we incorporate probability and random walks;
to discuss steady-state diffusion and laminar fluid flow, we use simple differential equations,
that we find solutions for by direct substitution.
Our module on mathematical modeling leads us to
simple linear algebra. The incorporation
of eigenvalues and eigenvectors in IPLS might initially seem ambitious,
but, in fact, the technical demands on the students
are that they be able to take a derivative of an exponential function of time,
and that they then be able to solve two simultaneous equations in two variables
to find the eigenvalues and eigenvectors, both of which they are able to do.
On the other hand,  multi-dimensional integrals are largely avoided,
removing a source of significant discomfort.
 
At the end of the first semester,  every year we have carried out an anonymous survey
to ascertain students' opinions.
We offer students credit towards their final grade for completing the survey and the response rate every year exceeds 95\%.
When asked about the level of mathematics,
it was ``way too advanced for my current ability'' for just 2\% of the students,
``advanced but manageable'' for 23\%,  ``about equal to my current ability'' for 46\%,
``below my current ability'' for 22\%, and ``well-below my current ability'' for 7\%
(Table~\ref{LevelOfMath}).
Concerning the level of calculus specifically,
it is ``too high'' for  19\%, ``just right'' for  66\%, and ``too low'' for 15\%
(Table~\ref{LevelOfCalculus}).
In view of these responses, we judge that the mathematics we use is appropriate.

\begin{table}[t!]
\center
\begin{tabular}{|c|c|c|c|c|}
\hline
Well below&Below my&Equal to&Advanced but&Way too\\
my current&current&my current&manageable&advanced\\
ability&ability&ability&manageable&advanced\\
\hline
0.07&0.22&0.46&0.23&0.02\\
\hline
\end{tabular}
\caption{Fraction of multiple-choice responses
to the question "Select the statement that best represents your opinion
concerning the level of math used."
}
\label{LevelOfMath}
\end{table}

\begin{table}[b!]
\center
\begin{tabular}{|c|c|c|}
\hline
Too low&Just right&Too high\\
\hline
0.15&0.66&0.19\\
\hline
\end{tabular}
\caption{Fraction of multiple-choice responses
to the question "What is your opinion
concerning the level of calculus used?"
}
\label{LevelOfCalculus}
\end{table}

Of course, because of varying student interests, preparations, and
motivations,  there cannot and should not be
a one-size-fits-all IPLS course,
suitable for every situation.
However, I believe that there is a place for a course like PHYS 170 at many institutions.
As biological science faculty introduce new upper-level biology electives
and major tracks
that demand mathematical sophistication,
in quantitative and systems biology, for example,
or as they incorporate more mathematical elements into
upper-level classes in ecology, epidemiology, neuroscience,
and evolutionary biology,
a physics class similar to PHYS 170 will
naturally serves as an essential prerequisite.

\section{Student reception}
How is PHYS 170 received by the students?
Significant numbers of students appreciate the fresh and powerful perspective
that physics brings to the life sciences,
 as the following quotes by anonymous students
 show:



\begin{quote}
``Thank you for class this semester. It has been really great to make links between all of my science courses at Yale, and in many ways (and I am shy to admit this, but against my expectations) PHYS 170 provided the platform for just that.''
\end{quote}




\begin{quote}
``This class is amazing if you genuinely like biology. If you're a biology major because you're premed or whatever you might not like it as much, but if you really care about biology this class is great. Physics is the future of biology, and this class gives you a taste of all the cool ways we can use quantitative techniques to describe living systems.
It definitely doesn't teach to the MCAT, go spend your summer savings on a Kaplan course if you want that. That doesn't exist at Yale and shouldn't.''
\end{quote}

\begin{quote}
``The biological aspects of this class really opened my eyes and changed my perspective of the world! You'll def learn some really fascinating and awesome stuff if you genuinely like biology or chemistry, and it was incredibly cool to approach chemical/biological processes from the different perspective of this class."
\end{quote}


\begin{quote}
``I LOVE the topics presented in 170. It really ties in to what I've learned in my other science classes (biology, biochemistry, and to a lesser extent chemistry), and it's wonderful to gain a new perspective/insight into mechanisms that I've just taken for granted, like diffusion, actin filament polymerization, enzyme kinetics, etc. It's just a really cool class, and has made me realize how much I love science - to the extent that it has, in part, motivated me to double major in another science field.''
\end{quote}

Although we emphasize at the start of the semester that PHYS 170
does not follow a traditional syllabus, and although Yale offers a parallel class
that follows a traditional syllabus at a comparable
level of mathematics, a number of students  nevertheless indicate that they wish that
PHYS 170 followed a traditional syllabus,
or
express concern that the class will not properly prepare them for the
Medical College Admissions Test (MCAT):

\begin{quote}
``I would have preferred to maybe learn more fundamental physics problems, such as those concerning a ball rolling down a hill. Though objectively more boring, I feel that those types of problems would give me a more solid physics foundation.''
\end{quote}

\begin{quote}
``I don't think I have gained a strong understanding of general physics concepts that I should know, especially since I plan on taking the MCAT in the future. I do not feel prepared at all.''
\end{quote}

\begin{quote}
``I cannot stand the probability unit. I understand why and how it is applicable to physics and biology and pre-med, but this is not a statistics course. I did not intend or want to learn these topics, but rather wished to learn what would be tested on the MCAT - classical physics.''
\end{quote}
However, a new version of the MCAT was introduced at the beginning of 2015.
Now, Kaplan, which provides MCAT preparation materials, advertises that
``physics will only be tested in the context of biological systems'',
and that momentum, periodic motion (springs and pendulums), circular motion,
center of mass, and the elastic properties of solids are excluded \cite{Kaplan}.
The best current advice for the instructors of introductory physics classes,
namely Ref.~\onlinecite{Hilborn2014}, concludes that
 \begin{quote}
`` .... life science students would be helped
  if their introductory physics courses
 included more work on transport processes
 (diffusion, osmosis, and fluid flow),
 on statistical reasoning (stochastic models),
 and microscopic models of materials ......''
 \end{quote}
 similar to a number of the actual choices for PHYS 170.

Lickert-type scale feedback, from our anonymous surveys,
 provides an overview of how the students, who have taken PHYS 170,
 view aspects of the course, related to pace, work load, and required time commitment.
Students mainly consider that the pace of the class  is ``just right"  (40\%) or "fast" (53\%)
(Table \ref{Pace}), and that
 the level of difficulty is ``about what I expected'' (42\%) or ``harder'' (35\%) (Table \ref{Harder}).
 They consider that the amount of
 time they spend on physics is "just right" (39\%) or "too much" (44\%)
 (Table~\ref{AmountOfTime}), but the actual length of time 85\%
 of students spend on physics is 12 hours or less (Table~\ref{HowManyHours}),
 which we judge to be reasonable.
These responses indicate that students generally view PHYS 170 as challenging class that requires considerable effort, but they do not view it as a class that requires excessive effort.

\begin{table}[t]
\center
\begin{tabular}{|c|c|c|c|c|}
\hline
A snail's pace&Slow&Just right&Fast&Lightening fast\\
\hline
0.00&0.02&0.40&0.53&0.06\\
\hline
\end{tabular}
\caption{Fraction of multiple-choice responses
to the question "How would you assess the pace of physics?".
}
\label{Pace}
\end{table}

\begin{table}[t]
\center
\begin{tabular}{|c|c|c|c|c|}
\hline
Much easier&Easier&About what I expected&Harder&Much harder\\
\hline
0.02&0.09&0.42&0.35&0.12\\
\hline
\end{tabular}
\caption{Fraction of multiple-choice responses
to the question "Has physics been a harder course, about the same or easier than you imagined it would be?".
}
\label{Harder}
\end{table}

\begin{table}[t]
\center
\begin{tabular}{|c|c|c|c|c|}
\hline
Way too little&Too little&Just right&Too much&Way too much\\
\hline
0.005&0.04&0.39&0.44&0.13\\
\hline
\end{tabular}
\caption{Fraction of multiple-choice responses
to the question "What is your feeling regrading the amount of time you spend on physics?".
}
\label{AmountOfTime}
\end{table}

\begin{table}[t]
\center
\begin{tabular}{|c|c|c|c|c|}
\hline
0-6 hours&6-9 hours&9-12 hours&12-15 hours&$>$15 hours\\
\hline
0.25&0.39&0.23&0.10&0.03\\
\hline
\end{tabular}
\caption{Fraction of multiple-choice responses
to the question "How many hours do you spend on physics outside of class?".
}
\label{HowManyHours}
\end{table}

PHYS 170 is one of four introductory physics courses offered at Yale,
including a course at a similar overall mathematical level
that follows a traditional syllabus, and
 two higher-level courses aimed at students with progressively  stronger
 mathematics and physics preparations.
To compare student opinions concerning
PHYS 170 to student opinions
concerning Yale's
traditional introductory physics courses,
we have tracked student evaluations of teaching (SETs)
for the nine years from 2006 to 2014 (Table~\ref{SETMatrix}).
Although the appropriateness of using
SETs as a measure of faculty
teaching effectiveness continues to be questioned \cite{Stark2014,Boring2016},
here,
we seek to use SETs as a means to gain insight into what students think about
different flavors of introductory physics.

Yale's SETs ask for numerical responses to two questions,
that facilitate comparisons among classes:
(1) Overall, how would you rate the work load
of this course in comparison to other Yale courses you have taken?
(Scale: 1=much less, 2=less, 3=same, 4=greater, 5=much greater);
and
(2) What is your overall assessment of this course? (Scale: 1=poor, 2=below average, 3=good, 4=very good, 5=excellent).
The responses to these questions from the previous three years are then made available to students
considering a class.

 \begin{figure}[t!]
\centering
{\includegraphics[width=0.49\textwidth,keepaspectratio=true]{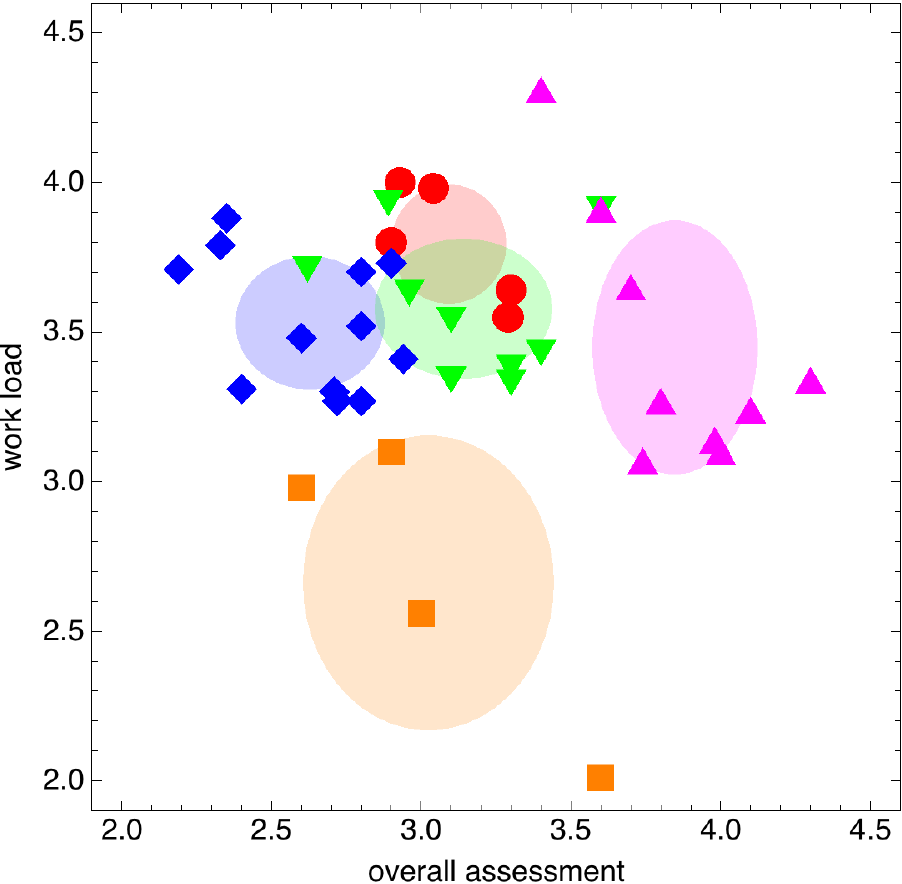}}
\caption{
Scatter plot of mean ``overall assessment'' and mean ``work load'' from student evaluations of
Yale's first-semester introductory physics classes over the last nine years.
Shown as orange squares, red circles, blue diamonds, green inverted triangles and magenta
triangles are the evaluations of PHYS 150,
PHYS 170, PHYS 180, PHYS 200, and PHYS 260, respectively.
}
\label{TeachingEvaluations}
\end{figure}

\begin{table*}[t]
\small
\center
\begin{tabular}{|c||c|c|c|c|c|c|c|c|c|c|c|}
\hline
Year&2006&2007&2008&2009&2010&2011&2012&2013&2014&Mean&Standard Deviation\\
\hline
\hline
PHYS 150 overall rating&2.9&3.0&3.6&2.6&&&&&&3.03&0.42\\
\hline
PHYS 150 work load&3.10&2.56&2.01&2.98&&&&&&2.66&0.49\\
\hline
PHYS 150 instructor&1&1&1&2&&&&&&&\\
\hline
\hline
PHYS 170 overall rating&&&&&2.9&3.3&3.29&2.98& 3.06&3.09&0.19\\
\hline
PHYS 170 work load&&&&&3.8&3.64&3.55&4.00&3.98&3.79& 0.20\\
\hline
PHYS 170 instructor&&&&&3&3&3&3&3&&\\
\hline
PHYS 170 enrollment&&&&&97&82&148&143&127&&\\
\hline
\hline
PHYS 180 overall rating&2.6&2.8&2.9&2.8&2.4&2.8&2.71, 2.35&2.72, 2.94& 2.19, 2.33&2.63&0.25\\
\hline
PHYS 180 work load&3.48&3.70&3.73&3.27&3.31&3.52&3.33, 3.88&3.27, 3.41&3.71, 3.79&3.53&0.22\\
\hline
PHYS 180 instructor&4&5&5, 6&7&7&8&8, 9&8, 10&8, 10&&\\
\hline
\hline
PHYS 200 overall rating&3.6&3.3&3.1&3.3&3.4&3.1&2.62&2.89& 2.96&3.14&0.30\\
\hline
PHYS 200 work load&3.6&3.38&3.34&3.33&3.43&3.54&3.71&3.93&3.63&3.58&0.23\\
\hline
PHYS 200 instructor&11&12&12&13&12&12&14&14&15&&\\
\hline
\hline
PHYS 260 overall rating&3.6&3.4&3.7&4.3&4.1&4.0&3.98& 3.80&3.74&3.85&0.28\\
\hline
PHYS 260 work load&3.91&4.31&3.65&3.34&3.24&3.10&3.14&3.27&3.07&3.45&0.42\\
\hline
PHYS 260 instructor&16&17&18&18&18&18&18&18&18&&\\
\hline
\hline
\end{tabular}
\caption{Student evaluations of teaching for introductory physics classes at Yale from 2006 through 2014.
The two numbers listed for the overall rating and work load of PHYS 180
for 2012, 2013, and 2014 are because this class was then offered in two sections.
Different instructors are labelled 1 through 18.
}
\label{SETMatrix}
\end{table*}

Fig. ~\ref{TeachingEvaluations}
is a scatter plot of the mean (averaged over student respondents)
of each these two ratings for each class of
first-semester introductory physics,
offered at Yale over the last nine years.
Shown as the red circles are the evaluations for
PHYS 170.
Shown as the orange squares are the evaluations for PHYS 150 {\em General Physics},
which was a traditional algebra-based physics class, aimed at pre-medical students, and a subset of Yale's biological science students, whose majors did not require calculus-based physics.
PHYS 170 was introduced in Fall 2010,
as a direct replacement for PHYS 150.
Shown as the blue diamonds are the evaluations for
PHYS 180 {\em University Physics}, which is calculus-based introductory physics,
at a similar overall mathematical level to PHYS 170,
that follows a traditional syllabus.
The audience for PHYS 180 includes most engineering majors, as well as a
significant number of premedical students.
The green inverted triangles show the evaluations for PHYS 200 {\em Fundamentals of Physics},
which is aimed at students with a strong interest in physical sciences, and
which covers Newtonian mechanics, including gravitation and rotational dynamics,
thermodynamics, and wave motion, at a more mathematically-advanced level
than PHYS 180.
Finally, the magenta triangles show
the evaluations for PHYS 260 {\em Intensive Introductory Physics},
which treats topics beyond the AP Physics syllabus, and
which is aimed at especially precocious students,
who already possess
a strong physics and mathematics background, and
who already have completed a year-long calculus-based
AP Physics course in mechanics and E\&M, and a course in vector calculus.
Physics-majors-to-be usually take either PHYS 200 or PHYS 260.

Evidently, the evaluations for each flavor of Yale's first-semester introductory physics
 appear to form more-or-less distinct clusters.
This point of view is emphasized by the large ellipses,
also shown in Fig. ~\ref{TeachingEvaluations}.
The center of each ellipse represents the means (averaged over years and sections)
of the ratings of each class, while the semi-major and semi-minor axes of the ellipse
represent the corresponding standard deviations.
The statistical significance of these groupings can be assessed quantitatively by
calculating $p$-values,
namely the probabilities that the SETs of pairs of classes are at least as divergent as observed,
assuming the null hypothesis that the SETs of both classes are drawn
from the same probability distribution.
Table~\ref{PvalueMatrix} presents the $p$-values for both overall rating
and work load for each pair of classes.
At least one of the two $p$-values for each pair of classes is less than 0.05 (except for the PHYS 170/PHYS 200 pair) demonstrating that the differences between classes are indeed statistically significant,
according to the convention that a $p$-value of less than 0.05 indicates statistical significance.
Because, over the nine-year period considered, each of these classes has been taught
in a number of different styles ({\em e.g.} traditional lecture or flipped classroom,
small or large class size) and/or by a number of different instructors,
it is 
plausible (but not proven) that the student evaluations
depend more on the audience and class content (subject matter and level of difficulty)
than on the particular instructor or teaching style.

\begin{table*}[t]
\footnotesize
\center
\begin{tabular}{|c||c|c|c|c|c|}
\hline
&PHYS 170&PHYS 180&PHYS 200&PHYS 260\\
\hline
\hline
PHYS150&0.81, 0.016, 0.22 ,3.2&0.14, 0.0037, 1.3, 2.6&0.52, 0.0028, 0.35, 2.8&0.042, 0.042, 2.6, 1.8\\
\hline
PHYS170&&0.0035, 0.086, 2.0, 1.6&0.87, 0.23, 0.18, 0.97&0.0010, 0.086, 3.0, 0.95\\
\hline
PHYS180&&&0.0049, 0.33, 1.8, 0.47&0.000041, 0.32, 4.7, 0.087\\
\hline
PHYS200&&&&0.0063, 0.11, 2.5, 0.38\\
\hline
\end{tabular}
\caption{Statistical significance and size of the differences
among SETs for Yale's introductory physics classes.
The first number of each quadruplet is the $p$-value, specifying
the probability that the overall ratings of the pair of classes in question are at least as divergent as
observed, assuming the null hypothesis that they are drawn from the same probability distribution.
The second number corresponds to the $p$-value for work loads of the pair.
These $p$-values are
calculated using the Kolmogorov-Smirnov test.
Conventionally, a $p$-value less than 0.05 is taken to imply that the difference between the two distributions being
compared is statistically significant.
The third number is the value of Cohen's $d$ for the overall ratings of the pair, which is a measure of how large is the
difference in their overall ratings. Specifically,
Cohen's $d$ is given by
$d = {|\mu_1-\mu_2 |}/{\sqrt{\frac{(n_1-1)\sigma_1^2 +(n_2-1) \sigma_2^2}{n_1+n_2-2}}}$,
where $n_1$ and $n_2$, $\mu_1$ and $\mu_1$, and $\sigma_1$ and $\sigma_2$ are the
number of samples, the means, and the standard deviations, respectively, of the two distributions being compared.
The fourth number is the value of Cohen's $d$ for the work loads of the pair.
}
\label{PvalueMatrix}
\end{table*}

Clearly, the overall evaluation of PHYS 260 is about three standard deviations higher than the
overall evaluations of PHYS 150, PHYS 170, and PHYS 200,
all of which have similar overall evaluations.
(Cohen's $d=2.6$, 3.0, and 2.5 for the difference between overall evaluations of PHYS 150, PHYS 170, and PHYS 200,
respectively, and PHYS 260.) 
In turn,  the overall evaluations of PHYS 150, PHYS 170, and PHYS 200
are nearly two standard deviations larger than the overall evaluation of PHYS 180.
(Cohen's $d=1.3$, 2.0, and 1.8 for the difference between the overall evaluations of PHYS 150, PHYS 170, and PHYS 200,
respectively, and PHYS 180.) 
At the same time, the work load of PHYS 150 was more than two standard deviations below the
work loads of PHYS 180, PHYS 200, and PHYS 260, which are all similar to each other,
while the work load of PHYS 170 is about one
standard deviation greater than those of these three classes.
(Cohen's $d= 2.6$, 2.8, and 1.8 for the difference between the work loads of PHYS 180, PHYS 200 and PHYS 260, respectively,
and PHYS 150, while
Cohen's $d=1.6$, 0.97, and 0.95  for
difference in
the work loads of PHYS 180,
PHYS 200, and PHYS 260, respectively, and PHYS 170).

Interestingly, within the evaluations for both PHYS 260 and PHYS 150,
for both of which there is a relatively broad range of numerical values for the mean
work load and the mean overall rating, there is clearly a strong negative correlation.
between
these two measures: the smaller the work load, the higher the overall rating, and {\em vice versa}.
There is also a negative correlation between work load and overall assessment within
the evaluations for PHYS 170 
although the range of both is smaller in this case than
 for either of PHYS 150 or PHYS 260.
This same trend can be seen in the ratings of PHYS 150 and
PHYS 180 in comparison to each other,
both of which follow a traditional introductory physics syllabus,
and both of which
enroll(ed) a preponderance of students,
taking physics because it is a compulsory requirement of their major,
and/or because is a requirement for admission to medical school.

It is not unexpected that
 the very well-prepared physics-majors-to-be,
 who learn about quantum mechanics, relativity, and other compelling topics in PHYS 260,
 rate that class considerably more highly
 than the biology majors and pre-medical students rate
 PHYS 150 and PHYS 180, where they
 learn about two-dimensional projectile motion and cylinders rolling
down inclined planes.
What is striking from these data, however, is that when this same population of biology majors
and pre-medical students is involved in a class,
namely PHYS 170, which is
built on a selection of biologically
and medically relevant topics,
even when the work load is high,
 these students then award an overall assessment rating that is comparable to
 the rating that this same population awards
a traditional-syllabus class with a much smaller work load
namely PHYS 150, and that a population of physical-science or engineering majors-to-be
gives to a traditional-syllabus physics class, namely PHYS 200.

 \begin{figure*}[t!]
\centering
{\includegraphics[width=0.7\textwidth,keepaspectratio=true]{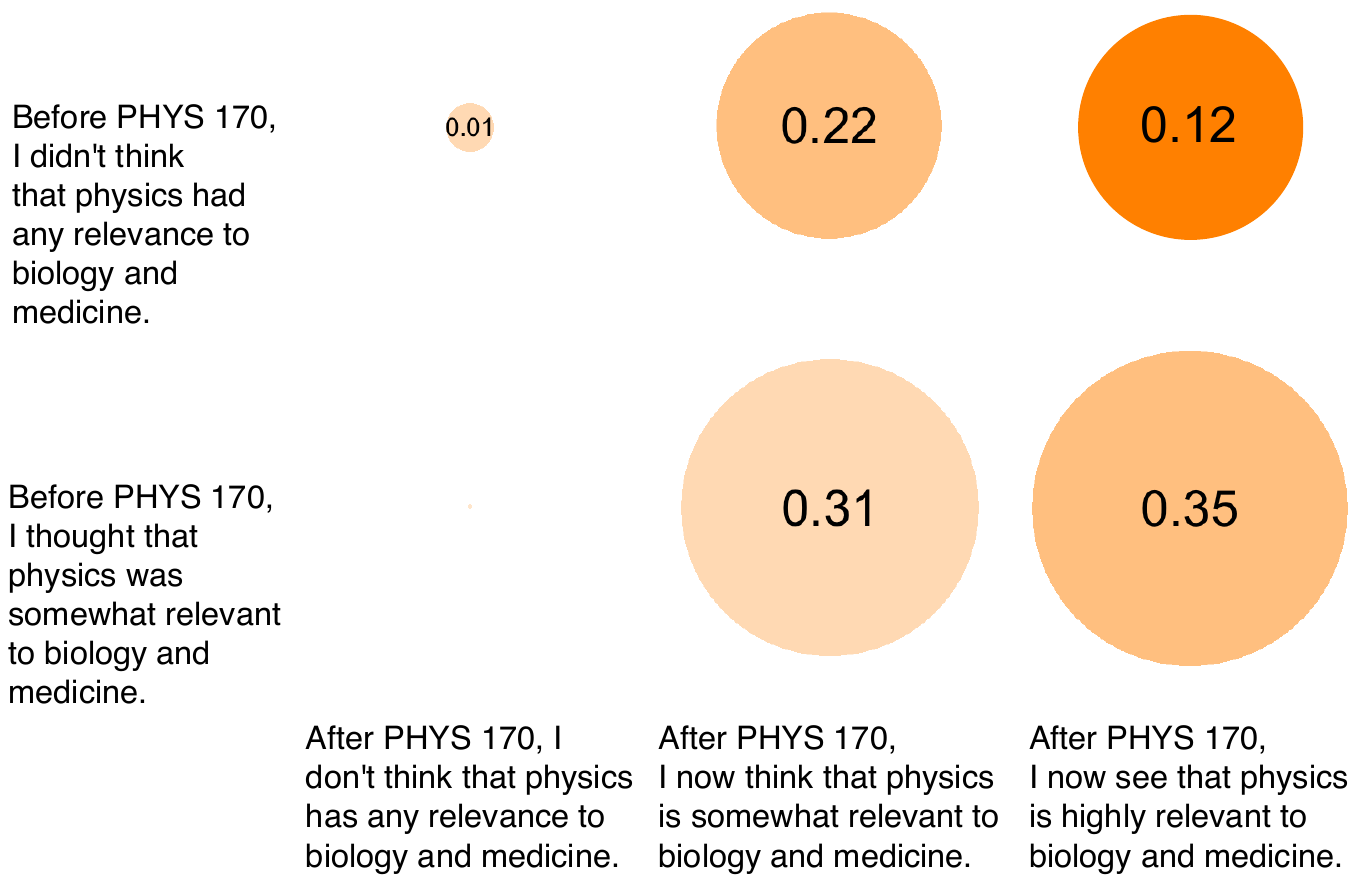}}
\caption{
Student opinions concerning the relevance of physics to biology and medicine.
The fraction of students responding in a particular way,
specified within a circle whose area is proportional to the fraction in question.
The response rate was 97\%.
}
\label{Feedback}
\end{figure*} 

Finally,  Fig.~\ref{Feedback}
displays PHYS 170 students' responses
when asked about how taking PHYS 170 had affected their opinion
concerning the relevance of physics to biology and medicine,
which was added to our survey in 2014 (124 responses out of 127 students).
Gratifyingly, nearly 70\% of these students report that taking the class lead to an increased
belief in the relevance of physics to biology and medicine:
12\%  and 22\% report that, prior to taking PHYS 170, they thought that physics had no relevance to
biology and medicine, but that afterwards they believe that physics is, respectively, highly
relevant  or somewhat relevant to biology and medicine;
35\% report that prior to taking PHYS 170 they considered physics to have some relevance to
biology and medicine, but that after they consider physics to be highly relevant to biology and
medicine. After having taken PHYS 170,
only 1\% of students persist with an initial opinion that physics has no relevance
to biology and medicine.

\section{Conclusions}

In comparison to traditional first-semester introductory physics,
biological science majors and premedical
students give a comparable or higher overall rating (PHYS 150 or PHYS 180, respectively)
to a version of first-semester introductory
physics (PHYS 170) that significantly adjusts the introductory physics syllabus to incorporate
a selection of
topics that are much more biologically and medically relevant,
even though their work load is considerably
or somewhat increased (PHYS 150 or PHYS 180, respectively).
Subsequent to PHYS 170,  these students show markedly increased belief that physics is relevant to biology and medicine.

The number one challenge in teaching PHYS 170 by far is students' lack of facility with algebraic manipulations.
Therefore, the first few weeks of PHYS 170 are, in part,
a crash algebra refresher for a significant number of its students.
It is difficult to see how to address this problem at the departmental level.
A related challenge, that emerged early on in the implementation, was complaints from
physics colleagues that PHYS 170 material was too difficult for
biology majors and premedical students. Experience has shown
that this is not the case, and such complaints have now  dissipated.

\begin{acknowledgments}
I would like to express my thanks to the wonderful students who have taken
the class,
especially  Titi Afolabi, Monique Arnold, Chayma Boussayoud, Millie Chapman, Angela Chen, Betsy Cowell, Regina De Luna,
Bertie Geng,  Mansur Ghani,
Sabrina Gill, Caleb Huang,
Syed Hussaini,
Ragini Luthra,
Yamini Naidu,
Maria Passarelli,
Jessica Perfetto, Faten Syed, Linda Wang, Victor Wang, 
Christine Willinger, Zizi Yu, and Gazelle Zerafati,
many of whom returned
to serve as Peer Tutors for the next year's class.
I would like to also express my gratitude to the
fantastic Graduate Teaching Fellows that I have had the opportunity to work with:
Tonima Ananna,
Ross Boltyanskiy, Jane Cummings,
Elizabeth Boulton, Shany Danieli,  Stephen Eckel,
Stefan Ellrington,  Meredith Frey, Judith Hoeller, Eric Holland, 
Amber Jessop, Anna Kashkanova, Peter Koo,
Lawrence Lee, Andrew Mack, Catherine Matulis, Wambui Mutoru, Danielle Norcini,
Jay Patel, Susan Pratt,
Alexsander Rebane, Jared Rovny, Brooke Russell, Raphael Sarfati, Daniel Seara, 
Olivier Trottier, Lucie Tvrznikova,
Kyle Vander Werf, Gennady Voronov,  Qing Xia, Yao Zhao, and Yuqi Zhu.
I am also deeply indebted to  Sean Barrett,
Sarah Demers,
Eric Dufresne, Jennifer Frederick, Stephen Irons,
Corey O'Hern, 
Rona Ramos,
Nicholas Read, William Segraves, Paul Tipton, and especially Sidney Cahn,  Claudia De Grandi, and Lynne Regan for their
invaluable advice and support.
I also thank Michael Choma, Scott Holley, and Thomas Pollard for wonderful guest lectures.
Finally, I  thank NSF PHY 1522467 and the Raymond and Beverley Sacker Institute for Physical
and Engineering Biology
for support.
\end{acknowledgments}


\end{document}